\documentclass[aps,superscriptaddress,prl,twocolumn,footinbib]{revtex4-1}
\usepackage{amsmath,amssymb}
\usepackage{bm}
\usepackage{epsfig}
\usepackage{natbib}
\usepackage{hyperref}
\usepackage{graphicx}
\usepackage{epstopdf}

 \hypersetup{pdfstartview={FitH},pdfpagemode={UseNone},
            colorlinks,linkcolor=red, citecolor=blue, urlcolor=blue,
            bookmarks=true, bookmarksopen=true, pdfnewwindow=true}
             
\usepackage{verbatim} % for multiple-line comment
\usepackage{morefloats}
\usepackage{bibentry}
\usepackage[mathscr]{euscript}

\newcommand{\rmi}{{\rm i}}
\newcommand{\rmd}{{\rm d}}

\newcommand{\e}{{\rm e}}

\renewcommand{\phi}{\varphi}
\begin{document}

\title{Distinguishing trivial and topological   quadrupolar  insulators by Wannier-Stark ladders}
\begin{abstract}I study theoretically quadrupolar topological insulators under applied  static electric field rotated along the crystal axis. I demonstrate, that the energy spectrum of this structure is a  Wannier-Stark ladder that is  quantized and directly distinguishes between the topological phase, possessing  localized corner states, and the trivial phase, lacking the corner states. These results may find applications in the characterization of   rapidly emerging higher-order topological phases of light and matter.
\end{abstract}
\author{\firstname{Alexander N.} \surname{Poddubny}}
\email{poddubny@coherent.ioffe.ru}
\affiliation{Ioffe  Institute, St.~Petersburg 194021, Russia}

\maketitle

{\it Introduction.} Berry phase and Chern numbers are now well established  concepts to characterize the excitations in crystalline solids~\cite{bernevig2013}.
 Namely, depending on the value of the Chern numbers, calculated for bulk Bloch bands in  infinite crystals, the finite samples, made from the same materials, will or will not have topological states propagating along  their edges. 
 Various interference techniques have been proposed to measure  the Berry phase for Bloch bands, and  the correspondence between bulk and edge features has been tested experimentally~\cite{Atala2013,Mittal2016}.
 
Recently, a quadrupolar two-dimensional (2D) topological phase has been put forward~\cite{Benalcazar61}. Contrary to the traditional 2D topological insulators, the quadrupolar phase has localized corner states, rather than  propagating 1D edge states. Such corner states have been demonstrated experimentally in the microwave~\cite{Peterson2018},  electric \cite{Imhof2018} and optical~\cite{Mittal2018} setups. The topological nature of the structure is manifested by the fractional quantized corner charges and edge polarizations. The bulk topological feature is  not exactly the Berry phase but the quantized phase of so-called nested Wilson loops calculated for the Bloch bands~\cite{Benalcazar61,Benalcazar2017PRB}. 
The Wilson loops  are significantly more complex entities than the Berry connection due to their inherently non-Abelian nature. Hence, the traditional  interference techniques developed to measure the Berry phase  \cite{Atala2013,Poshakinskiy2015,Mittal2016} are inapplicable. The Wilson loop tomography has so far been developed  only for the case of non-degenerate Bloch bands~\cite{Li1094}, while the quadrupolar phase has double-degenerate Bloch bands due to the reflection and chiral symmetries~\cite{Benalcazar2017PRB}. 
Hence, the fundamental question of measurable bulk manifestations of the quadrupolar topological phase remains open and  a clear protocol to measure the topological invariants in the bulk is highly desired.

Here, I put forward a procedure for bulk spectroscopy of topological features of the quadrupolar phase based on the application of a constant electric field $\bm F$ to the structure. 
 The energy spectrum of the general one-dimensional biased periodic structure is a Wannier-Stark ladder of discrete levels~ \cite{Mendez1993,Shevchenko2010,Gluck2002}. The energy levels $E_{n}$ of the ladder formed from a given Bloch band depend
 on the electric field as $E_{n} =E_{n}(F=0)+F(n+c)$, where $n=0,\pm 1,\ldots $ is the level number \cite{Vanderbilt2014,Maksimov2015,Lee2015} and $c$ are the Wannier center positions \cite{Vanderbilt1997}.
  It is known for   a 1D Su-Schrieffer-Heeger  structure  \cite{Maksimov2015} as well as  for 2D and 3D topological insulators~ \cite{Vanderbilt2014,Lee2015,Kolovsky2018}, that the centers $c=\{ \rmd E_{n}/\rmd F_{x}\}$ contain information on the topological invariants ($\{x\}$ is the fractional part of $x$). Here,  I extend this concept to the  quadrupolar phase in a {\it rotated} electric field, $F_{x}\ne 0$, $F_{y}\ne 0$. I  demonstrate, that the  shifts $\{\rmd E/\rmd F_{y}\}$ are quantized for $F_{x}\ne 0$ and naturally reveal  the topological phases of nested Wilson loops. This  provides  a transparent connection between the formal mathematical definition of the topological quadrupolar phase and its physical manifestations observable in the bulk.
 %%%%%%%%%%%%%%%%%%%%%%%%%%%%%%%%%%%
\begin{figure}[b]
\centering\includegraphics[width=0.4\textwidth]{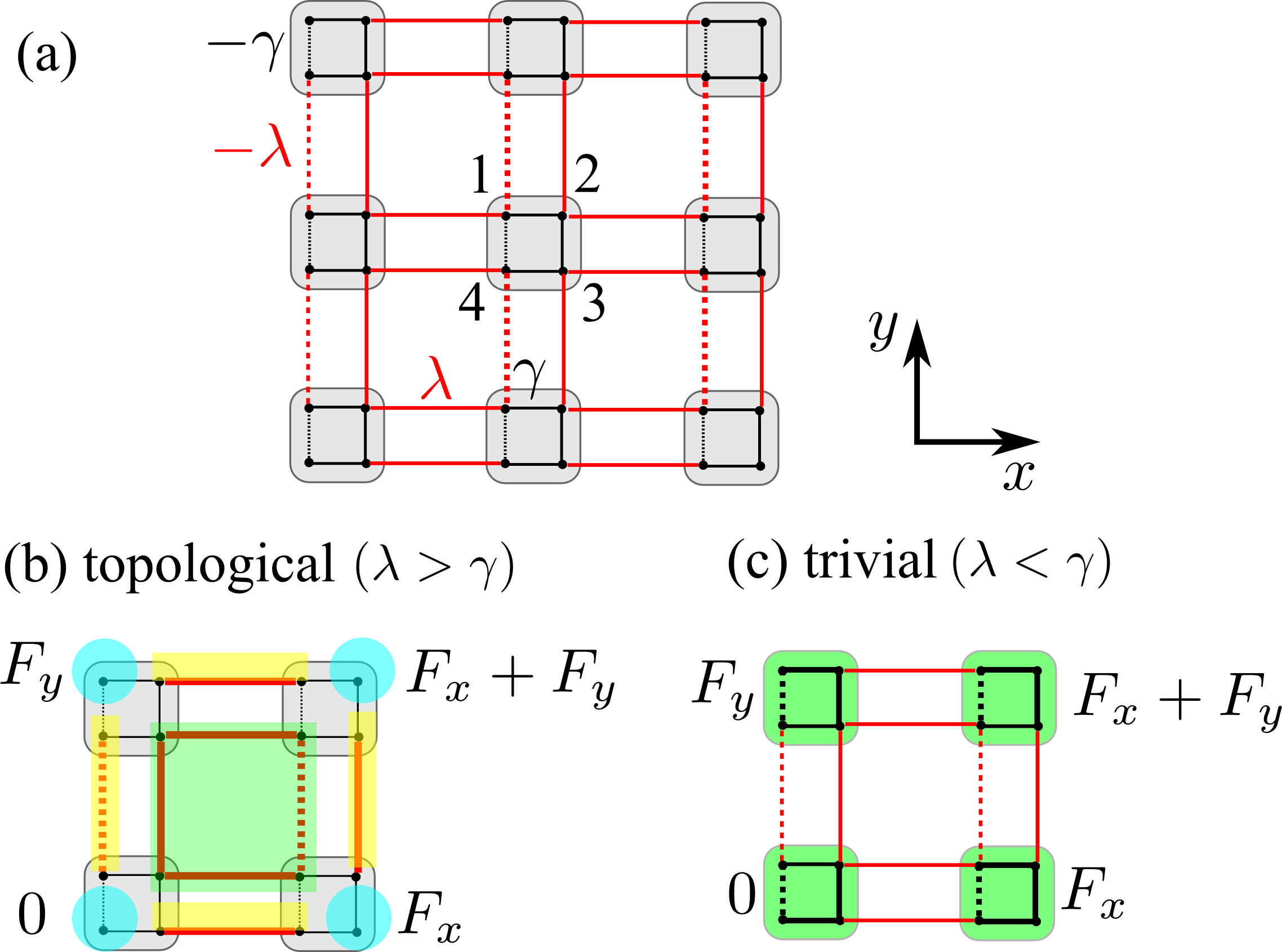}\\
\caption{Scheme of the structure under consideration. (a) Unbiased structure. (b,c) Biased topological (b) and trivial (c) structures. Green, blue and yellow shading depicts bulk, corner and edge states, respectively. External potential in the four unit cells is indicate on graph. }
\label{fig:1}
\end{figure}
%%%%%%%%%%%%%%%%%%%%%%%%%%%%%%%%%%%
 
{\it Quadrupolar phase in the rotated electric field.} The structure under consideration is schematically illustrated in Fig.~\ref{fig:1}(a)~\cite{Benalcazar2017PRB}. It can be described by a tight-binding Hamiltonian on a square lattice  with the alternating intra-cell and inter-cell tunneling constants $\gamma$ and $\lambda$ (black and red solid lines). Dotted lines correspond to the couplings $-\gamma$ and $-\lambda$. Alternation of the positive and negative couplings ensures the non-zero $\pi$-flux through the unit cell, and opens the band gap in the energy spectrum of the infinite structure~\cite{Benalcazar61}. The momentum space Hamiltonian for the periodic structure reads
\begin{align}\label{eq:H}
&H(k_{x},k_{y})=\\\nonumber&\begin{pmatrix}
0&\gamma+\lambda\e^{-\rmi k_{x}}&0&-\gamma-\lambda\e^{\rmi k_{y}}\\
\gamma+\lambda\e^{\rmi k_{x}}&0& \gamma+\lambda \e^{\rmi k_{y}}&0\\
0& \gamma+\lambda \e^{-\rmi k_{y}}&0&\gamma+\lambda \e^{\rmi k_{x}}\\
-\gamma-\lambda \e ^{-\rmi k_{y}} &0&\gamma +\lambda \e^{-\rmi k_{x}}&0
\end{pmatrix}\:,
\end{align}
where $\bm k$ is the Bloch wave vector and the basis corresponds to the atoms in the unit cell ordered as indicated in Fig.~\ref{fig:1}(a). Depending on the ratio between the couplings $\gamma$ and $\lambda$, the structure can be either in topological, or in a trivial phase. Namely, for $|\gamma|<|\lambda|$ [Fig.~\ref{fig:1}(b)] it has corner states (blue circles) and edge states (yellow lines) in addition to the bulk states (green shading). In the opposite case,  $|\gamma|>|\lambda|$ [Fig.~\ref{fig:1}(c)] the corner and edge states are absent.  This is seemingly  similar to the case of 1D  SSH model, where the termination with a weak tunneling link leads to the formation of a zero-energy edge state~\cite{bernevig2013}. However, the quadrupolar phase is qualitatively different from the 2D SSH  lattice because of the presence of negative couplings that are essential for the formation of bulk band gap and topologically protected corner states,  absent in a 2D SSH  system~\cite{Mittal2018}. 

The goal of this work is to distinguish between the topological and trivial phases from  the bulk, rather than edge, spectral features by applying the external electric field $
\bm F$. The main idea is sketched in Fig.~\ref{fig:1}(b,c). I assume that all  four sites in each unit cell with the discrete coordinates $x,y=0,\pm 1,\pm 2\ldots$ are biased by the energy  $F_{x}x+F_{y}y$.  
 In the trivial phase with $\gamma\gg \lambda$, [Fig.~\ref{fig:1}(c)], each Wannier  state is fully located inside the given unit cell. The states have two-fold degeneracy and the energies $\pm \sqrt{2}\gamma$. The external potential does not split these states and  only shifts them independently. In the topological phase with strong intra-cell couplings, $ \lambda\gg \gamma$ the centers of the Wannier functions (green squares) are located between the unit cells~\cite{Benalcazar2017PRB}, see Fig.~\ref{fig:1}(b). As such, each Wannier state is shared between four unit cells with different values of the external potential, and, contrary to the trivial case, the spectrum of the Wannier states will be split by the potential. Next, I prove this crude analysis by the rigorous calculation of the Wannier-Stark ladder.

%%%%%%%%%%%%%%%%%%%%%%%%%%%%%%%%%%%
\begin{figure}[t]
\includegraphics[width=0.45\textwidth]{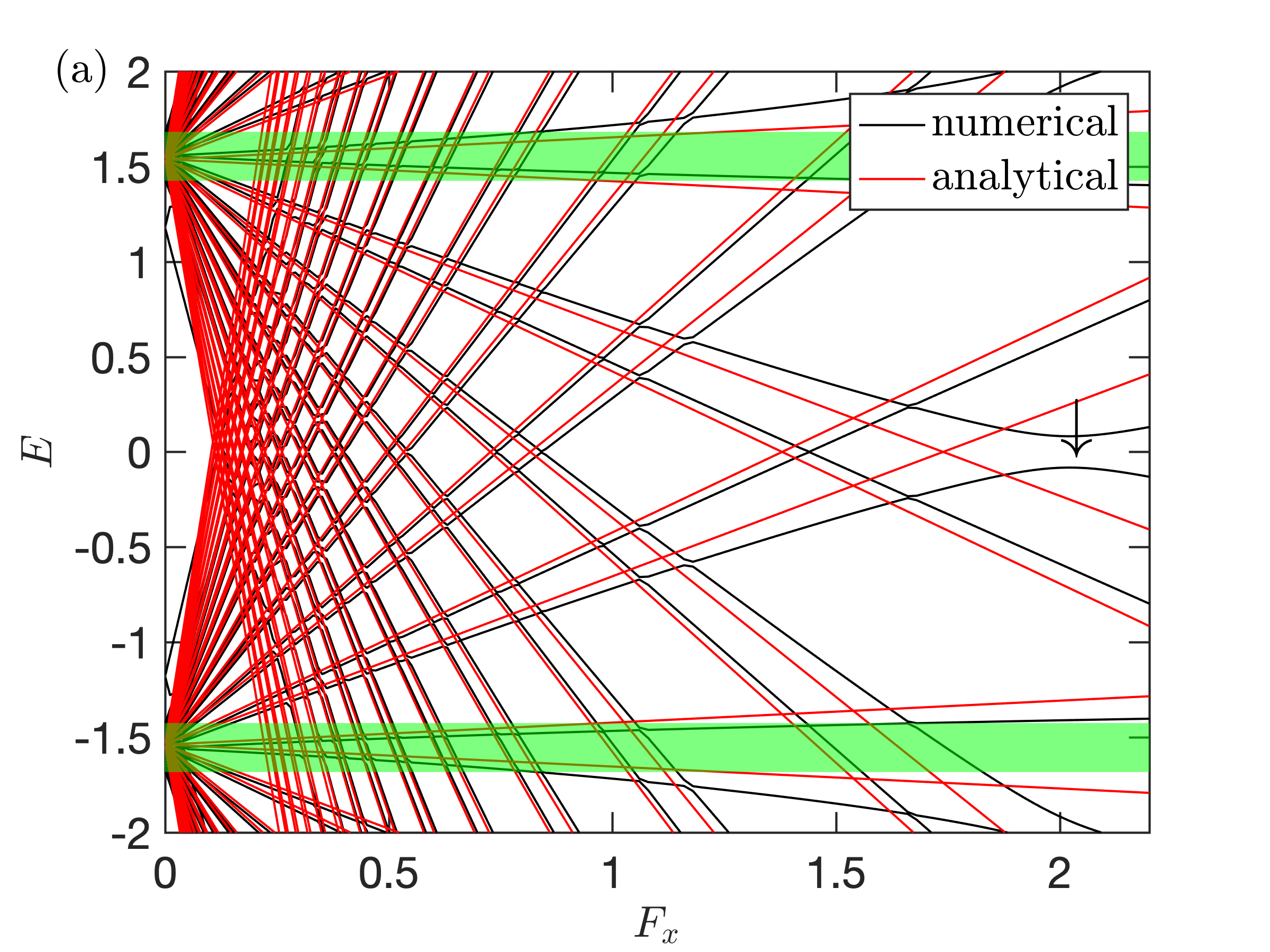}\\
\includegraphics[width=0.45\textwidth]{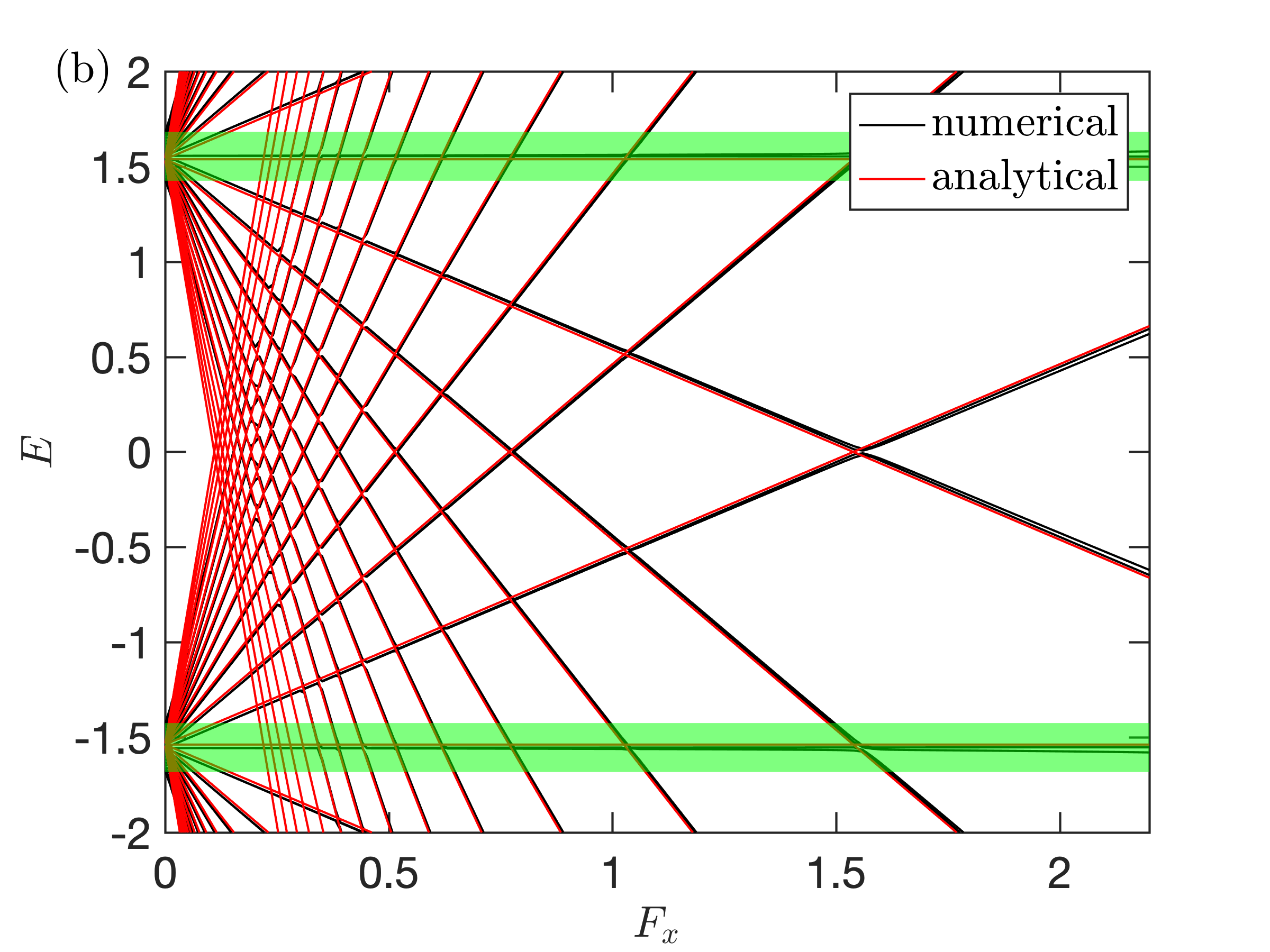}
\caption{Stark ladders depending on $F_{x}$  calculated for $F_{y}=0$. Panels (a) and (b) correspond to the topological ($\gamma=0.2$, $\lambda=1$) and trivial ($\gamma=1$, $\lambda=0.2$) structure. Black lines have been calculated numerically and red lines correspond to the analytical results Eq.~\eqref{eq:stark1}, Eq.~\eqref{eq:stark2}. Green shading shows  the Bloch bands in the unbiased periodic structure.  Vertical arrow in (a) indicates the splitting due to the Landau-Zener effect. Periodic boundary conditions with $k_{y}=0.5$ have been used  along the $y$ direction,  open boundary conditions with $N=10$ unit cells have been used  along $x$.
}\label{fig:1DStark}
\end{figure}
%%%%%%%%%%%%%%%%%%%%%%%%%%%%%%%%%%%
%%%%%%%%%%%%%%%%%%%%%%%%%%%%%%%%%%%
\begin{figure}[t]
\centering\includegraphics[width=0.45\textwidth]{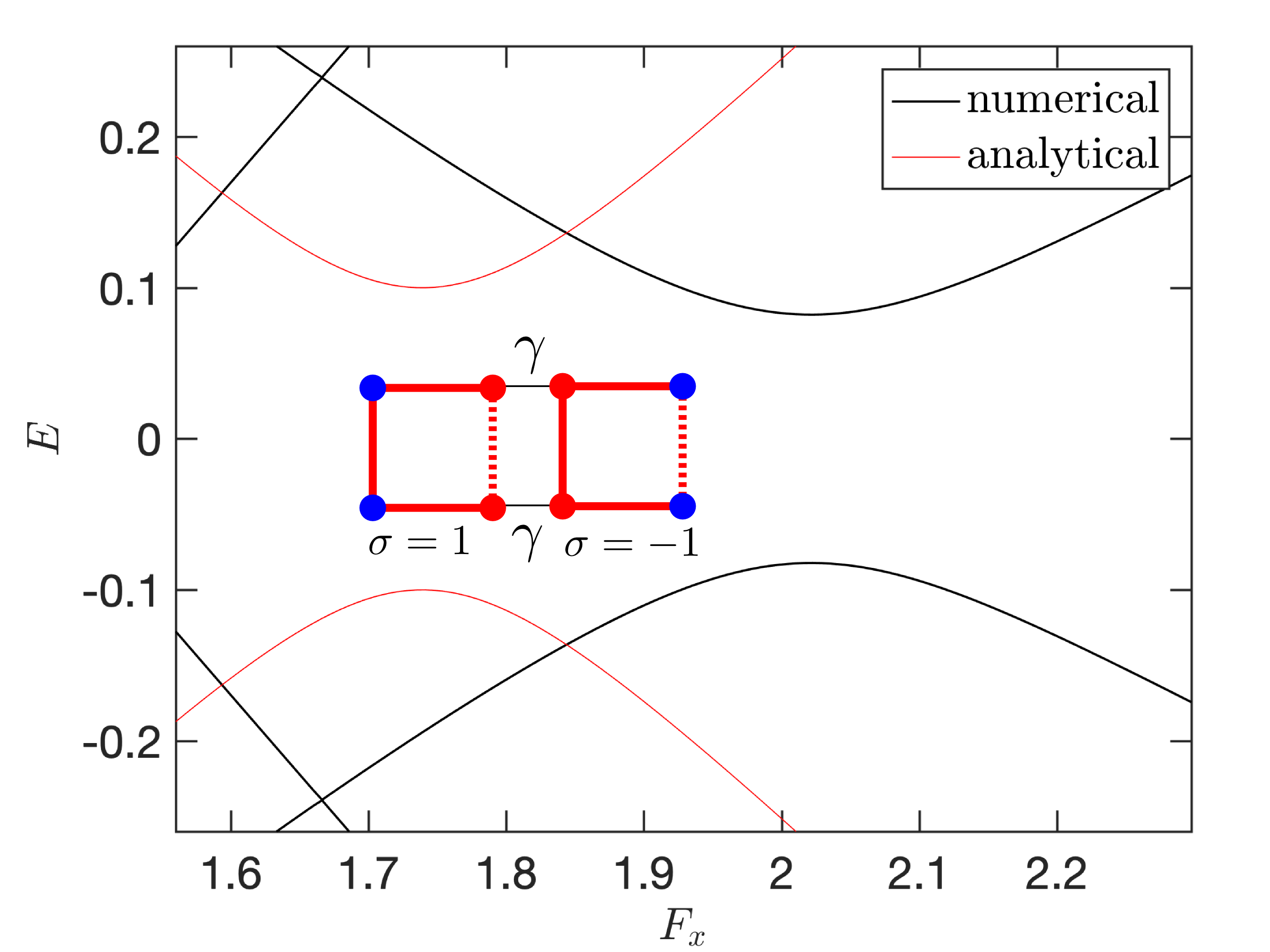}
\caption{ Spectrum of the Wannier-Stark ladder from Fig.~\ref{fig:1DStark}(a) enlarged in the vicinity of $F_{x}=2$. Black lines show the results of numerical calculation, red lines show the analytical result Eq.~\eqref{eq:LZ}.
Inset schematically illustrates the spatial structure of the  two Wannier functions of the coupled  states, with red/blue circles corresponding to the values of $\psi(x,y)=\pm 1/2$.}
\label{fig:LZ}
\end{figure}
%%%%%%%%%%%%%%%%%%%%%%%%%%%%%%%%%%%
%%%%%%%%%%%%%%%%%%%%%%%%%
{\it Stark ladders for $F_{y}=0$}.  I start by considering the structure with the electric field applied along the $x$ direction and the periodic  boundary conditions with the wave vector $k_{y}$ along $y$.  The energy spectra in the topological and trivial phases are shown in Fig.~\ref{fig:1DStark}(a) and Fig.~\ref{fig:1DStark}(b), respectively. The calculation demonstrates that the bulk Bloch bands are split due to the electric field and the two fans of levels, with the energies linear in electric field, emerge from each band. In the limit of $\gamma\ll \lambda$ ($\lambda\ll \gamma$) the energy levels are approximately given by the analytical expressions
\begin{align}
E_{n\sigma}^{(\rm topo)}&\approx \sigma \left(\sqrt{2}\lambda+\frac{\gamma\cos k_{y}}{\sqrt{2}}\right)+F_{x}\left(n+\frac{1}{2}\pm \frac{\sqrt{2}}{4}q\right),\label{eq:stark1}\\
E_{n\sigma}^{(\rm triv)}&\approx \sigma\left(\sqrt{2}\gamma+\frac{\lambda\cos k_{y}}{\sqrt{2}}\right)+nF_{x}\:,\label{eq:stark2}
\end{align}
where $n=0,\pm 1\ldots$, $\sigma=\pm 1$ distinguishes the upper and lower bands and $q=1+\gamma\cos k_{y}/(2\lambda)$.  These equations  can be rigorously obtained  as $E_{n\sigma}=F_{x}(n+c_{\pm})$, where $c_{\pm}=
\ln(\lambda_{\pm})/2\pi\rmi$ and $\lambda$ are the eigenvalues of the $2\times 2$ Wilson loop operator
$W=\exp\{\rmi/F_{x} \int_{0}^{2\pi}\rmd k_{x}[E_{\sigma}(k_{x})+ \rmi F_{x}u_{\sigma}^{\dag } \partial u_{\sigma}/\partial k_{x}]\}$. Here
$E_{\sigma}=\sigma \sqrt{2}\sqrt{\lambda^{2}+\gamma^{2}+\lambda\gamma(\cos k_{x}+\cos k_{y})}$ are the Bloch band energies and the $4\times 2$ matrix $u_{\sigma}=[u_{1\sigma},u_{2\sigma}]$ contains two  Bloch functions $u_{1}$ and $u_{2}$ for the corresponding band $\sigma$~\footnote{See Supplemental Materials for the  details}. The analytical answers Eqs.~\eqref{eq:stark1},\eqref{eq:stark2} are shown by the red lines in Fig.~\ref{fig:1DStark} and well describe the  numerical calculation. The difference between the topological and trivial cases is now clearly seen. In agreement with the naive explanation in Fig.~\ref{fig:1}(b,c), in the topological case the levels Eq.~\eqref{eq:stark1} are split by $\approx \sqrt{2} F_{x}/2$, while in the trivial case the states Eq.~\eqref{eq:stark2} remain almost twice degenerate.

%%%%%%%%%%%%%%%%%%%%%%%%%%%%%%%%%%%
{\it Landau-Zener effect.} 
Equations \eqref{eq:stark1},\eqref{eq:stark2} describe  two independent Stark ladders formed from the  two Bloch bands.  However, it can be seen from Fig.~\ref{fig:1DStark}(a), that in fact these ladders are not independent and feature avoided crossings. The largest anticrossing  in Fig.~\ref{fig:1DStark}(a) takes place at $F_{x}\approx 2$ and is indicated by an arrow. In Fig.~\ref{fig:LZ} I have plotted the same levels in the larger scale around the field $F_{x}\approx 2$. Its physical origin can be understood by analyzing the two Wannier functions of the upper and lower bands ($\sigma=\pm 1$). Specifically, in the limit of $\lambda\gg \gamma$ the Wannier functions are fully localized on 4 sites between the unit cells [see Fig.~\ref{fig:1}(b)] with the absolute values of  all four nonzero wavefunction components  equal to $1/2$. The parameter $n$  in Eq.~\eqref{eq:stark1} specifies the horizontal coordinate, i.e. the state $n$ is located between the $n$-th and $n+1$-th unit cells. Hence, for $F_{x}\approx 1.7\lambda$ one has $E_{1,-}\approx E_{0,+}$  which means that the  Wannier states of upper and lower bands, shifted by one site, become degenerate (see the inset of Fig.~\ref{fig:LZ}). The tunneling between these states lifts the degeneracy and opens the gap with the width $\gamma$. Indeed, the analytical result \cite{Note1}
%%%%%%%%%%%%%%%%%%%%%%%%%%%%%%%%%%%
\begin{equation}\label{eq:LZ}
\epsilon_{\pm}=\pm\frac{1}{2}\sqrt{(E_{1,-}-E_{0,+})^{2}+\gamma^{2}}.
\end{equation}
shown by the red lines in Fig.~\ref{fig:LZ} well describes the exact numerical answer. The only difference is  the small horizontal shift of the position of the anticrossing, that is a next order effect in the parameter  $F/\lambda$, not taken into account in the approximation Eq.~\eqref{eq:stark1}, Eq.~\eqref{eq:LZ}. However, the width of the gap for the black curves in Fig.~\ref{fig:LZ} is close to $\gamma=0.2$ in agreement with Eq.~\eqref{eq:LZ}.
%%%%%%%%%%%%%%%%%%%%%%%%%%%%%%%%%%%
\begin{figure}[t]
\includegraphics[width=0.5\textwidth]{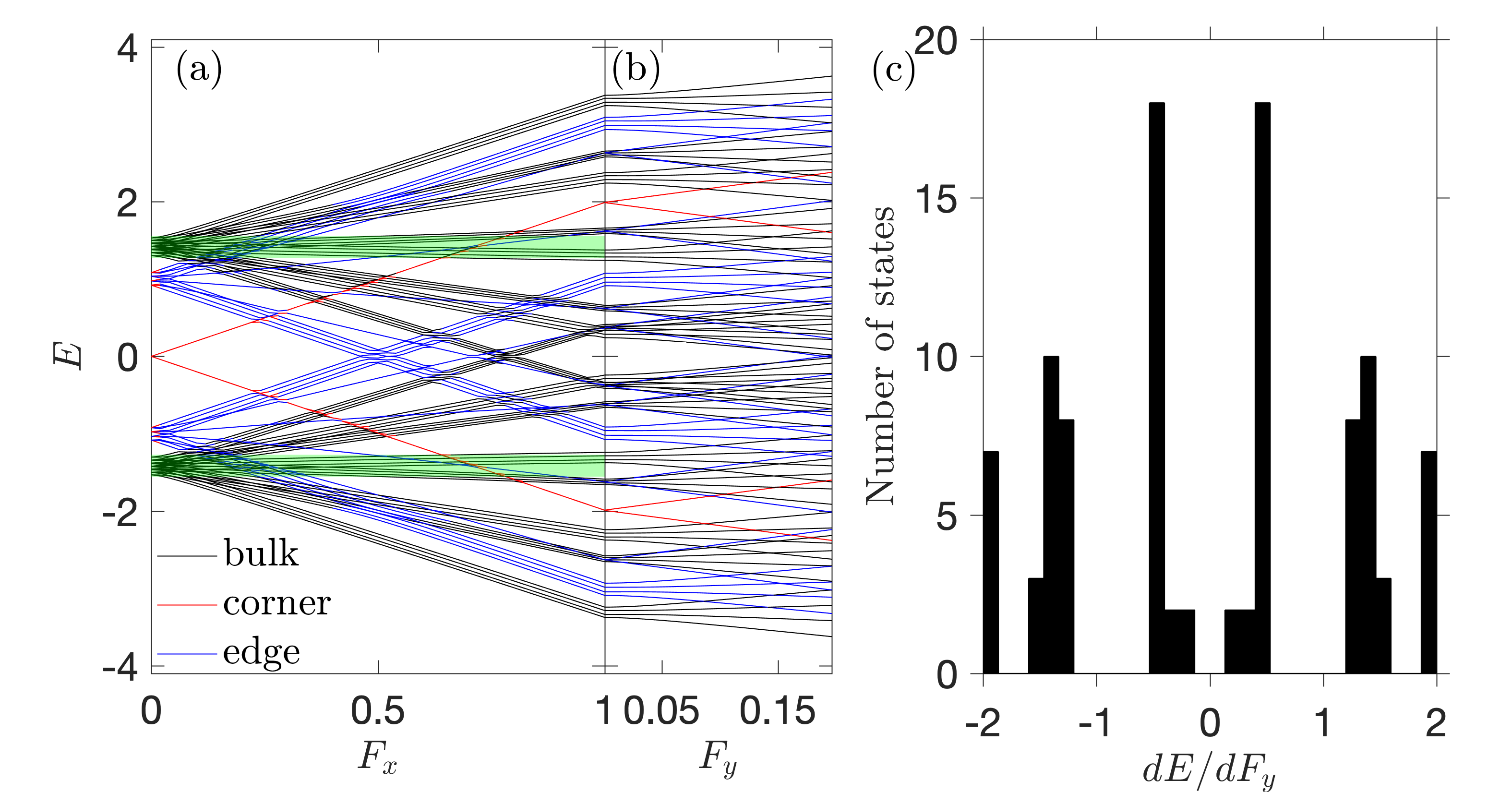}\\
\includegraphics[width=0.5\textwidth]{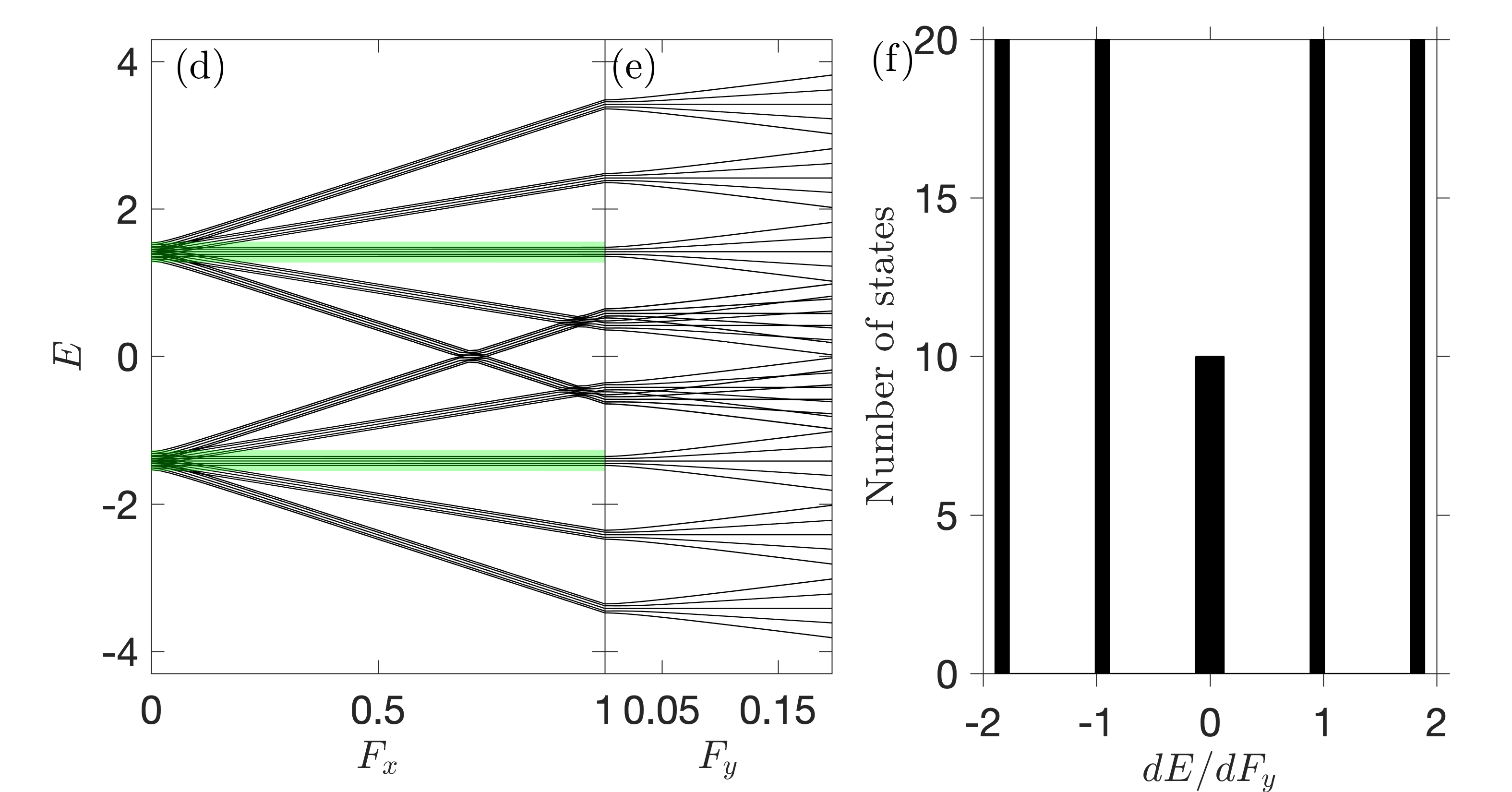}
\caption{Stark ladders for an open lattice with $5\times 5$ unit cells depending on $F_{x}$(a,d) and $F_{y}$ (b,e)
in the topological (a,b,c) and trivial (d,e,f) phases with $\gamma=0.15$, $\lambda=1$ and 
$\gamma=1$, $\lambda=0.15$, respectively. Panels (c,f) show the distributions  of $\rmd E/\rmd F_{y}$ for $F_{x}=1$, $F_{y}=0.1$. Green shading indicates the  Bloch bands in the unbiased periodic structure.  
 }\label{fig:nest}
\end{figure} 
%%%%%%%%%%%%%%%%%%%%%%%%%

{\it Nested Wilson loops via the Stark ladders.}
It has been demonstrated in Fig.~\ref{fig:1DStark}  that the application of the electric field modifies the structure in such way that  its eigenstates become the Wannier functions, depending on the wave vector $k_{y}$.  
The phase of the nested  Wilson loops proposed in Refs.~\cite{Benalcazar61,Benalcazar2017PRB} is just the  winding number of these functions calculated when $k_{y}$ is varied across the Brillouin zone. I have verified numerically that this phase  is equal to $\pi$ ($2\pi$) for the topological (trivial) states in Fig.~\ref{fig:1DStark}(a) [ Fig.~\ref{fig:1DStark}(b)] as expected~\cite{Note1}. Instead of the calculation of this phase under the periodic boundary conditions along the $y$ direction one can consider a more realistic situation of a  finite structure, open from all four sides. Application of  an additional electric field along the $y$ direction, corresponding to the rotation of the total electric field,  will then  further split the  states in Fig.~\ref{fig:1DStark}. Next, I will demonstrate by an explicit numerical calculation, that the resulting {\it nested Stark ladder} is quantized and directly reflects the phases of the nested Wilson loops.

%%%%%%%%%%%%%%%%%%%%%%%%%%%%%%%%%%%
\begin{figure}[t]
\includegraphics[width=0.45\textwidth]{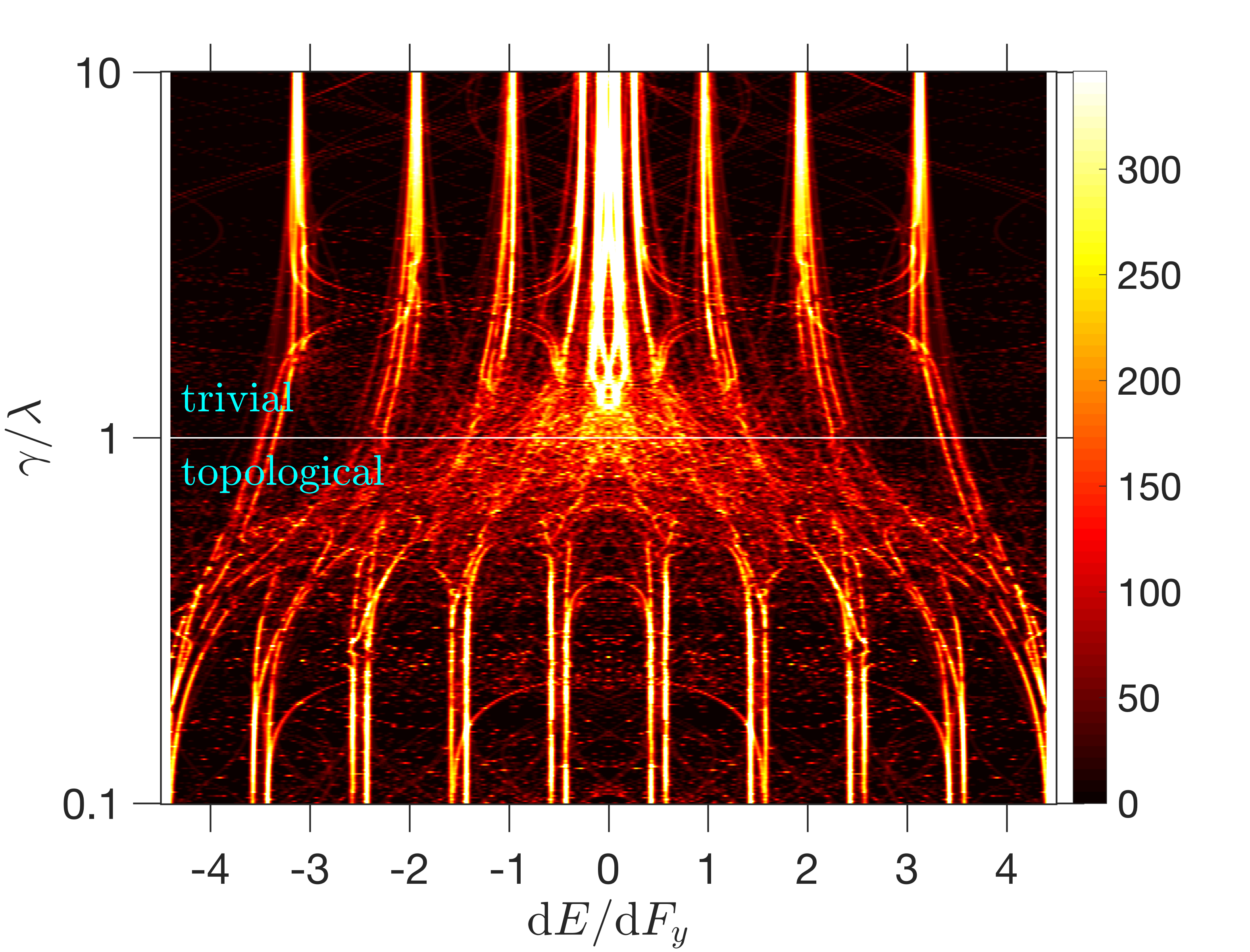}
\caption{Color map of the distribution $\rmd E/\rmd F_{y}$ depending on $\gamma$ for $\lambda=1$, $F_{x}=1$, $F_{y}=0.1$. Calculation has been performed for the structure with $13\times 13$ unit cells.}\label{fig:color}
\end{figure} 
%%%%%%%%%%%%%%%%%%%%%%%%%%%%%%%%%%%

The results of the calculation are presented in Fig.~\ref{fig:nest}. First, I apply the electric field along the $x$ direction, which leads to the splitting of the Bloch bands for both topological [Fig.~\ref{fig:nest}(a)]
and trivial [Fig.~\ref{fig:nest}(d)] structures, similarly to Fig.~\ref{fig:1DStark}. Since now the topological structure is open in both directions, it also has edge and corner modes, indicated by blue and red colors, respectively, in addition to the bulk modes. The edge (corner) states have been formally defined as the states where the probability of localization at the sites at the edge (corner sites) is larger than 0.2.
 Application of an additional electric field along the $y$ direction leads to the further splitting of the modes  [Fig.~\ref{fig:nest}(b,e)]. The key difference between the topological and trivial cases is the distribution of the derivatives $\rmd E/\rmd F_{y}$  for given $F_{x}\ne 0$, shown in Fig.~\ref{fig:nest}(d,f). In the topological structure this distribution has  maxima around the {\it half-integer} values $\pm 1/2,\pm 3/2\ldots$, i.e. $\{\rmd E/\rmd F_{y}\}=1/2$. In the trivial structure the distribution peaks are  at the {\it integer} values  $0,\pm 1,\pm 2\ldots$, i.e. $\{\rmd E/\rmd F_{y}\}=0$.  The broadening of the peaks, most distinct in the topological phase, is due to the finite size effects, that are relatively significant for the small structure with only $5\times 5$ unit cells. This distinction between half-integer and integer values of the derivatives exactly corresponds to the difference between the $\pi$ and $2\pi$ phase of the nested Wilson loops calculated in Ref.~\cite{Benalcazar61} for the topological and trivial structures. The advantage of the current construction with the Wannier-Stark ladders is that the topological phase arises naturally and can be accessible in finite open systems.

Finally, in Fig.~\ref{fig:color} I calculate the variation of the distribution $\rmd E/\rmd F_{y}$ for given fixed values of $F_{x}$ and $F_{y}$ depending on the ratio $\gamma/\lambda$.  Due to the larger size of the structure, more peaks are resolved as compared to Fig.~\ref{fig:nest}. In order to simplify the presentation, the peak distribution has been convoluted with a Gaussian function with the dispersion of $0.02$.
In agreement with the results in Fig.~\ref{fig:nest}, the positions of the peaks shift from half-integer to integer values
with the increase of $\gamma$. This reflects the transition from the topological to the trivial regime.

%%%%%%%%%%%%%%%%%%%%%%%%%%%%%%%%%%%
To summarize, I have  introduced the concept of bulk spectroscopy of topological invariants in the quadrupolar insulators by relating the quantized phases of nested Wilson loops to the spectra of Wannier-Stark ladders. The easiest experimental implementation of the current scheme might be provided by topoelectric circuits~\cite{Imhof2018}, where 
the on-site energies of individual sites can be tuned by introducing nonlinear elements~\cite{Khanikaev2018}.
These scheme can be directly generalized to other higher-order topological insulators~\cite{Schindler2018}. Another potential avenue for future research, complementary to the study of discrete energy levels, is the analysis of the Bloch oscillations ~\cite{Dmitriev2002,Lee2015} in the quadrupolar phase under rotated electric field.
%%%%%%%%%%%%%%%%%%%%%%%%%%%%%%%%%%%
\begin{acknowledgements}
 
I  acknowledge the financial support by the Foundation for Advancement of Theoretical Physics and Mathematics ``Basis'' and the Russian Foundation for Basic Research Grants No. 18-29-20037 and 18-32-20065.
\end{acknowledgements}

%\nocite{apsrev41Control}
%\bibliographystyle{apsrev4}
%\bibliography{titleon,quad}

\begin{thebibliography}{22}%
\makeatletter
\providecommand \@ifxundefined [1]{%
 \@ifx{#1\undefined}
}%
\providecommand \@ifnum [1]{%
 \ifnum #1\expandafter \@firstoftwo
 \else \expandafter \@secondoftwo
 \fi
}%
\providecommand \@ifx [1]{%
 \ifx #1\expandafter \@firstoftwo
 \else \expandafter \@secondoftwo
 \fi
}%
\providecommand \natexlab [1]{#1}%
\providecommand \enquote  [1]{``#1''}%
\providecommand \bibnamefont  [1]{#1}%
\providecommand \bibfnamefont [1]{#1}%
\providecommand \citenamefont [1]{#1}%
\providecommand \href@noop [0]{\@secondoftwo}%
\providecommand \href [0]{\begingroup \@sanitize@url \@href}%
\providecommand \@href[1]{\@@startlink{#1}\@@href}%
\providecommand \@@href[1]{\endgroup#1\@@endlink}%
\providecommand \@sanitize@url [0]{\catcode `\\12\catcode `\$12\catcode
  `\&12\catcode `\#12\catcode `\^12\catcode `\_12\catcode `\%12\relax}%
\providecommand \@@startlink[1]{}%
\providecommand \@@endlink[0]{}%
\providecommand \url  [0]{\begingroup\@sanitize@url \@url }%
\providecommand \@url [1]{\endgroup\@href {#1}{\urlprefix }}%
\providecommand \urlprefix  [0]{URL }%
\providecommand \Eprint [0]{\href }%
\providecommand \doibase [0]{http://dx.doi.org/}%
\providecommand \selectlanguage [0]{\@gobble}%
\providecommand \bibinfo  [0]{\@secondoftwo}%
\providecommand \bibfield  [0]{\@secondoftwo}%
\providecommand \translation [1]{[#1]}%
\providecommand \BibitemOpen [0]{}%
\providecommand \bibitemStop [0]{}%
\providecommand \bibitemNoStop [0]{.\EOS\space}%
\providecommand \EOS [0]{\spacefactor3000\relax}%
\providecommand \BibitemShut  [1]{\csname bibitem#1\endcsname}%
\let\auto@bib@innerbib\@empty
%</preamble>
\bibitem [{\citenamefont {Bernevig}\ and\ \citenamefont
  {Hughes}(2013)}]{bernevig2013}%
  \BibitemOpen
  \bibfield  {author} {\bibinfo {author} {\bibfnamefont {B.}~\bibnamefont
  {Bernevig}}\ and\ \bibinfo {author} {\bibfnamefont {T.}~\bibnamefont
  {Hughes}},\ }\href@noop {} {\emph {\bibinfo {title} {Topological Insulators
  and Topological Superconductors}}}\ (\bibinfo  {publisher} {Princeton
  University Press},\ \bibinfo {year} {2013})\BibitemShut {NoStop}%
\bibitem [{\citenamefont {{Atala}}\ \emph {et~al.}(2013)\citenamefont
  {{Atala}}, \citenamefont {{Aidelsburger}}, \citenamefont {{Barreiro}},
  \citenamefont {{Abanin}}, \citenamefont {{Kitagawa}}, \citenamefont
  {{Demler}},\ and\ \citenamefont {{Bloch}}}]{Atala2013}%
  \BibitemOpen
  \bibfield  {author} {\bibinfo {author} {\bibfnamefont {M.}~\bibnamefont
  {{Atala}}}, \bibinfo {author} {\bibfnamefont {M.}~\bibnamefont
  {{Aidelsburger}}}, \bibinfo {author} {\bibfnamefont {J.~T.}\ \bibnamefont
  {{Barreiro}}}, \bibinfo {author} {\bibfnamefont {D.}~\bibnamefont
  {{Abanin}}}, \bibinfo {author} {\bibfnamefont {T.}~\bibnamefont
  {{Kitagawa}}}, \bibinfo {author} {\bibfnamefont {E.}~\bibnamefont
  {{Demler}}}, \ and\ \bibinfo {author} {\bibfnamefont {I.}~\bibnamefont
  {{Bloch}}},\ }\bibfield  {title} {\enquote {\bibinfo {title} {{Direct
  measurement of the Zak phase in topological Bloch bands}},}\ }\href {\doibase
  10.1038/nphys2790} {\bibfield  {journal} {\bibinfo  {journal} {Nature
  Physics}\ }\textbf {\bibinfo {volume} {9}},\ \bibinfo {pages} {795--800}
  (\bibinfo {year} {2013})}\BibitemShut {NoStop}%
\bibitem [{\citenamefont {Mittal}\ \emph {et~al.}(2016)\citenamefont {Mittal},
  \citenamefont {Ganeshan}, \citenamefont {Fan}, \citenamefont {Vaezi},\ and\
  \citenamefont {Hafezi}}]{Mittal2016}%
  \BibitemOpen
  \bibfield  {author} {\bibinfo {author} {\bibfnamefont {S.}~\bibnamefont
  {Mittal}}, \bibinfo {author} {\bibfnamefont {S.}~\bibnamefont {Ganeshan}},
  \bibinfo {author} {\bibfnamefont {J.}~\bibnamefont {Fan}}, \bibinfo {author}
  {\bibfnamefont {A.}~\bibnamefont {Vaezi}}, \ and\ \bibinfo {author}
  {\bibfnamefont {M.}~\bibnamefont {Hafezi}},\ }\bibfield  {title} {\enquote
  {\bibinfo {title} {Measurement of topological invariants in a 2{D} photonic
  system},}\ }\href {\doibase 10.1038/nphoton.2016.10} {\bibfield  {journal}
  {\bibinfo  {journal} {Nature Photonics}\ } (\bibinfo {year} {2016}),\
  10.1038/nphoton.2016.10}\BibitemShut {NoStop}%
\bibitem [{\citenamefont {Benalcazar}\ \emph
  {et~al.}(2017{\natexlab{a}})\citenamefont {Benalcazar}, \citenamefont
  {Bernevig},\ and\ \citenamefont {Hughes}}]{Benalcazar61}%
  \BibitemOpen
  \bibfield  {author} {\bibinfo {author} {\bibfnamefont {W.~A.}\ \bibnamefont
  {Benalcazar}}, \bibinfo {author} {\bibfnamefont {B.~A.}\ \bibnamefont
  {Bernevig}}, \ and\ \bibinfo {author} {\bibfnamefont {T.~L.}\ \bibnamefont
  {Hughes}},\ }\bibfield  {title} {\enquote {\bibinfo {title} {Quantized
  electric multipole insulators},}\ }\href {\doibase 10.1126/science.aah6442}
  {\bibfield  {journal} {\bibinfo  {journal} {Science}\ }\textbf {\bibinfo
  {volume} {357}},\ \bibinfo {pages} {61--66} (\bibinfo {year}
  {2017}{\natexlab{a}})}\BibitemShut {NoStop}%
\bibitem [{\citenamefont {Peterson}\ \emph {et~al.}(2018)\citenamefont
  {Peterson}, \citenamefont {Benalcazar}, \citenamefont {Hughes},\ and\
  \citenamefont {Bahl}}]{Peterson2018}%
  \BibitemOpen
  \bibfield  {author} {\bibinfo {author} {\bibfnamefont {C.~W.}\ \bibnamefont
  {Peterson}}, \bibinfo {author} {\bibfnamefont {W.~A.}\ \bibnamefont
  {Benalcazar}}, \bibinfo {author} {\bibfnamefont {T.~L.}\ \bibnamefont
  {Hughes}}, \ and\ \bibinfo {author} {\bibfnamefont {G.}~\bibnamefont
  {Bahl}},\ }\bibfield  {title} {\enquote {\bibinfo {title} {A quantized
  microwave quadrupole insulator with topologically protected corner states},}\
  }\href {\doibase 10.1038/nature25777} {\bibfield  {journal} {\bibinfo
  {journal} {Nature}\ }\textbf {\bibinfo {volume} {555}},\ \bibinfo {pages}
  {346--350} (\bibinfo {year} {2018})}\BibitemShut {NoStop}%
\bibitem [{\citenamefont {Imhof}\ \emph {et~al.}(2018)\citenamefont {Imhof},
  \citenamefont {Berger}, \citenamefont {Bayer}, \citenamefont {Brehm},
  \citenamefont {Molenkamp}, \citenamefont {Kiessling}, \citenamefont
  {Schindler}, \citenamefont {Lee}, \citenamefont {Greiter}, \citenamefont
  {Neupert},\ and\ \citenamefont {Thomale}}]{Imhof2018}%
  \BibitemOpen
  \bibfield  {author} {\bibinfo {author} {\bibfnamefont {S.}~\bibnamefont
  {Imhof}}, \bibinfo {author} {\bibfnamefont {C.}~\bibnamefont {Berger}},
  \bibinfo {author} {\bibfnamefont {F.}~\bibnamefont {Bayer}}, \bibinfo
  {author} {\bibfnamefont {J.}~\bibnamefont {Brehm}}, \bibinfo {author}
  {\bibfnamefont {L.~W.}\ \bibnamefont {Molenkamp}}, \bibinfo {author}
  {\bibfnamefont {T.}~\bibnamefont {Kiessling}}, \bibinfo {author}
  {\bibfnamefont {F.}~\bibnamefont {Schindler}}, \bibinfo {author}
  {\bibfnamefont {C.~H.}\ \bibnamefont {Lee}}, \bibinfo {author} {\bibfnamefont
  {M.}~\bibnamefont {Greiter}}, \bibinfo {author} {\bibfnamefont
  {T.}~\bibnamefont {Neupert}}, \ and\ \bibinfo {author} {\bibfnamefont
  {R.}~\bibnamefont {Thomale}},\ }\bibfield  {title} {\enquote {\bibinfo
  {title} {Topolectrical-circuit realization of topological corner modes},}\
  }\href {\doibase 10.1038/s41567-018-0246-1} {\bibfield  {journal} {\bibinfo
  {journal} {Nature Physics}\ }\textbf {\bibinfo {volume} {14}},\ \bibinfo
  {pages} {925--929} (\bibinfo {year} {2018})}\BibitemShut {NoStop}%
\bibitem [{\citenamefont {{Mittal}}\ \emph {et~al.}(2018)\citenamefont
  {{Mittal}}, \citenamefont {{Vikram Orre}}, \citenamefont {{Zhu}},
  \citenamefont {{Gorlach}}, \citenamefont {{Poddubny}},\ and\ \citenamefont
  {{Hafezi}}}]{Mittal2018}%
  \BibitemOpen
  \bibfield  {author} {\bibinfo {author} {\bibfnamefont {S.}~\bibnamefont
  {{Mittal}}}, \bibinfo {author} {\bibfnamefont {V.}~\bibnamefont {{Vikram
  Orre}}}, \bibinfo {author} {\bibfnamefont {G.}~\bibnamefont {{Zhu}}},
  \bibinfo {author} {\bibfnamefont {M.~A.}\ \bibnamefont {{Gorlach}}}, \bibinfo
  {author} {\bibfnamefont {A.}~\bibnamefont {{Poddubny}}}, \ and\ \bibinfo
  {author} {\bibfnamefont {M.}~\bibnamefont {{Hafezi}}},\ }\bibfield  {title}
  {\enquote {\bibinfo {title} {{Photonic quadrupole topological phases}},}\
  }\href@noop {} {\bibfield  {journal} {\bibinfo  {journal} {arXiv e-prints}\ }
  (\bibinfo {year} {2018})}\BibitemShut {NoStop}%
\bibitem [{\citenamefont {Benalcazar}\ \emph
  {et~al.}(2017{\natexlab{b}})\citenamefont {Benalcazar}, \citenamefont
  {Bernevig},\ and\ \citenamefont {Hughes}}]{Benalcazar2017PRB}%
  \BibitemOpen
  \bibfield  {author} {\bibinfo {author} {\bibfnamefont {W.~A.}\ \bibnamefont
  {Benalcazar}}, \bibinfo {author} {\bibfnamefont {B.~A.}\ \bibnamefont
  {Bernevig}}, \ and\ \bibinfo {author} {\bibfnamefont {T.~L.}\ \bibnamefont
  {Hughes}},\ }\bibfield  {title} {\enquote {\bibinfo {title} {Electric
  multipole moments, topological multipole moment pumping, and chiral hinge
  states in crystalline insulators},}\ }\href {\doibase
  10.1103/PhysRevB.96.245115} {\bibfield  {journal} {\bibinfo  {journal} {Phys.
  Rev. B}\ }\textbf {\bibinfo {volume} {96}},\ \bibinfo {pages} {245115}
  (\bibinfo {year} {2017}{\natexlab{b}})}\BibitemShut {NoStop}%
\bibitem [{\citenamefont {Poshakinskiy}\ \emph {et~al.}(2015)\citenamefont
  {Poshakinskiy}, \citenamefont {Poddubny},\ and\ \citenamefont
  {Hafezi}}]{Poshakinskiy2015}%
  \BibitemOpen
  \bibfield  {author} {\bibinfo {author} {\bibfnamefont {A.~V.}\ \bibnamefont
  {Poshakinskiy}}, \bibinfo {author} {\bibfnamefont {A.~N.}\ \bibnamefont
  {Poddubny}}, \ and\ \bibinfo {author} {\bibfnamefont {M.}~\bibnamefont
  {Hafezi}},\ }\bibfield  {title} {\enquote {\bibinfo {title} {Phase
  spectroscopy of topological invariants in photonic crystals},}\ }\href
  {\doibase 10.1103/PhysRevA.91.043830} {\bibfield  {journal} {\bibinfo
  {journal} {Phys. Rev. A}\ }\textbf {\bibinfo {volume} {91}},\ \bibinfo
  {pages} {043830} (\bibinfo {year} {2015})}\BibitemShut {NoStop}%
\bibitem [{\citenamefont {Li}\ \emph {et~al.}(2016)\citenamefont {Li},
  \citenamefont {Duca}, \citenamefont {Reitter}, \citenamefont {Grusdt},
  \citenamefont {Demler}, \citenamefont {Endres}, \citenamefont
  {Schleier-Smith}, \citenamefont {Bloch},\ and\ \citenamefont
  {Schneider}}]{Li1094}%
  \BibitemOpen
  \bibfield  {author} {\bibinfo {author} {\bibfnamefont {T.}~\bibnamefont
  {Li}}, \bibinfo {author} {\bibfnamefont {L.}~\bibnamefont {Duca}}, \bibinfo
  {author} {\bibfnamefont {M.}~\bibnamefont {Reitter}}, \bibinfo {author}
  {\bibfnamefont {F.}~\bibnamefont {Grusdt}}, \bibinfo {author} {\bibfnamefont
  {E.}~\bibnamefont {Demler}}, \bibinfo {author} {\bibfnamefont
  {M.}~\bibnamefont {Endres}}, \bibinfo {author} {\bibfnamefont
  {M.}~\bibnamefont {Schleier-Smith}}, \bibinfo {author} {\bibfnamefont
  {I.}~\bibnamefont {Bloch}}, \ and\ \bibinfo {author} {\bibfnamefont
  {U.}~\bibnamefont {Schneider}},\ }\bibfield  {title} {\enquote {\bibinfo
  {title} {Bloch state tomography using {W}ilson lines},}\ }\href {\doibase
  10.1126/science.aad5812} {\bibfield  {journal} {\bibinfo  {journal}
  {Science}\ }\textbf {\bibinfo {volume} {352}},\ \bibinfo {pages} {1094--1097}
  (\bibinfo {year} {2016})}\BibitemShut {NoStop}%
\bibitem [{\citenamefont {Mendez}\ and\ \citenamefont
  {Bastard}(1993)}]{Mendez1993}%
  \BibitemOpen
  \bibfield  {author} {\bibinfo {author} {\bibfnamefont {E.~E.}\ \bibnamefont
  {Mendez}}\ and\ \bibinfo {author} {\bibfnamefont {G.}~\bibnamefont
  {Bastard}},\ }\bibfield  {title} {\enquote {\bibinfo {title} {Wannier-{S}tark
  ladders and {B}loch oscillations in superlattices},}\ }\href {\doibase
  10.1063/1.881353} {\bibfield  {journal} {\bibinfo  {journal} {Physics Today}\
  }\textbf {\bibinfo {volume} {46}},\ \bibinfo {pages} {34--42} (\bibinfo
  {year} {1993})}\BibitemShut {NoStop}%
\bibitem [{\citenamefont {Shevchenko}\ \emph {et~al.}(2010)\citenamefont
  {Shevchenko}, \citenamefont {Ashhab},\ and\ \citenamefont
  {Nori}}]{Shevchenko2010}%
  \BibitemOpen
  \bibfield  {author} {\bibinfo {author} {\bibfnamefont {S.}~\bibnamefont
  {Shevchenko}}, \bibinfo {author} {\bibfnamefont {S.}~\bibnamefont {Ashhab}},
  \ and\ \bibinfo {author} {\bibfnamefont {F.}~\bibnamefont {Nori}},\
  }\bibfield  {title} {\enquote {\bibinfo {title}
  {Landau{\textendash}{Z}ener{\textendash}{S}t\"{u}ckelberg interferometry},}\
  }\href {\doibase 10.1016/j.physrep.2010.03.002} {\bibfield  {journal}
  {\bibinfo  {journal} {Physics Reports}\ }\textbf {\bibinfo {volume} {492}},\
  \bibinfo {pages} {1--30} (\bibinfo {year} {2010})}\BibitemShut {NoStop}%
\bibitem [{\citenamefont {{Gl{\"u}ck}}\ \emph {et~al.}(2002)\citenamefont
  {{Gl{\"u}ck}}, \citenamefont {{R.~Kolovsky}},\ and\ \citenamefont
  {{Korsch}}}]{Gluck2002}%
  \BibitemOpen
  \bibfield  {author} {\bibinfo {author} {\bibfnamefont {M.}~\bibnamefont
  {{Gl{\"u}ck}}}, \bibinfo {author} {\bibfnamefont {A.}~\bibnamefont
  {{R.~Kolovsky}}}, \ and\ \bibinfo {author} {\bibfnamefont {H.~J.}\
  \bibnamefont {{Korsch}}},\ }\bibfield  {title} {\enquote {\bibinfo {title}
  {{Wannier-Stark resonances in optical and semiconductor superlattices}},}\
  }\href {\doibase 10.1016/S0370-1573(02)00142-4} {\bibfield  {journal}
  {\bibinfo  {journal} {Phys. Rep.}\ }\textbf {\bibinfo {volume} {366}},\
  \bibinfo {pages} {103--182} (\bibinfo {year} {2002})}\BibitemShut {NoStop}%
\bibitem [{\citenamefont {Taherinejad}\ \emph {et~al.}(2014)\citenamefont
  {Taherinejad}, \citenamefont {Garrity},\ and\ \citenamefont
  {Vanderbilt}}]{Vanderbilt2014}%
  \BibitemOpen
  \bibfield  {author} {\bibinfo {author} {\bibfnamefont {M.}~\bibnamefont
  {Taherinejad}}, \bibinfo {author} {\bibfnamefont {K.~F.}\ \bibnamefont
  {Garrity}}, \ and\ \bibinfo {author} {\bibfnamefont {D.}~\bibnamefont
  {Vanderbilt}},\ }\bibfield  {title} {\enquote {\bibinfo {title} {Wannier
  center sheets in topological insulators},}\ }\href {\doibase
  10.1103/PhysRevB.89.115102} {\bibfield  {journal} {\bibinfo  {journal} {Phys.
  Rev. B}\ }\textbf {\bibinfo {volume} {89}},\ \bibinfo {pages} {115102}
  (\bibinfo {year} {2014})}\BibitemShut {NoStop}%
\bibitem [{\citenamefont {Maksimov}\ \emph {et~al.}(2015)\citenamefont
  {Maksimov}, \citenamefont {Bulgakov},\ and\ \citenamefont
  {Kolovsky}}]{Maksimov2015}%
  \BibitemOpen
  \bibfield  {author} {\bibinfo {author} {\bibfnamefont {D.~N.}\ \bibnamefont
  {Maksimov}}, \bibinfo {author} {\bibfnamefont {E.~N.}\ \bibnamefont
  {Bulgakov}}, \ and\ \bibinfo {author} {\bibfnamefont {A.~R.}\ \bibnamefont
  {Kolovsky}},\ }\bibfield  {title} {\enquote {\bibinfo {title}
  {Wannier-{S}tark states in double-periodic lattices. {I}. {O}ne-dimensional
  lattices},}\ }\href {\doibase 10.1103/PhysRevA.91.053631} {\bibfield
  {journal} {\bibinfo  {journal} {Phys. Rev. A}\ }\textbf {\bibinfo {volume}
  {91}},\ \bibinfo {pages} {053631} (\bibinfo {year} {2015})}\BibitemShut
  {NoStop}%
\bibitem [{\citenamefont {Lee}\ and\ \citenamefont {Park}(2015)}]{Lee2015}%
  \BibitemOpen
  \bibfield  {author} {\bibinfo {author} {\bibfnamefont {W.-R.}\ \bibnamefont
  {Lee}}\ and\ \bibinfo {author} {\bibfnamefont {K.}~\bibnamefont {Park}},\
  }\bibfield  {title} {\enquote {\bibinfo {title} {Direct manifestation of
  topological order in the winding number of the wannier-stark ladder},}\
  }\href {\doibase 10.1103/PhysRevB.92.195144} {\bibfield  {journal} {\bibinfo
  {journal} {Phys. Rev. B}\ }\textbf {\bibinfo {volume} {92}},\ \bibinfo
  {pages} {195144} (\bibinfo {year} {2015})}\BibitemShut {NoStop}%
\bibitem [{\citenamefont {Marzari}\ and\ \citenamefont
  {Vanderbilt}(1997)}]{Vanderbilt1997}%
  \BibitemOpen
  \bibfield  {author} {\bibinfo {author} {\bibfnamefont {N.}~\bibnamefont
  {Marzari}}\ and\ \bibinfo {author} {\bibfnamefont {D.}~\bibnamefont
  {Vanderbilt}},\ }\bibfield  {title} {\enquote {\bibinfo {title} {Maximally
  localized generalized {W}annier functions for composite energy bands},}\
  }\href {\doibase 10.1103/PhysRevB.56.12847} {\bibfield  {journal} {\bibinfo
  {journal} {Phys. Rev. B}\ }\textbf {\bibinfo {volume} {56}},\ \bibinfo
  {pages} {12847--12865} (\bibinfo {year} {1997})}\BibitemShut {NoStop}%
\bibitem [{\citenamefont {Kolovsky}(2018)}]{Kolovsky2018}%
  \BibitemOpen
  \bibfield  {author} {\bibinfo {author} {\bibfnamefont {A.~R.}\ \bibnamefont
  {Kolovsky}},\ }\bibfield  {title} {\enquote {\bibinfo {title} {Topological
  phase transitions in tilted optical lattices},}\ }\href {\doibase
  10.1103/PhysRevA.98.013603} {\bibfield  {journal} {\bibinfo  {journal} {Phys.
  Rev. A}\ }\textbf {\bibinfo {volume} {98}},\ \bibinfo {pages} {013603}
  (\bibinfo {year} {2018})}\BibitemShut {NoStop}%
\bibitem [{Note1()}]{Note1}%
  \BibitemOpen
  \bibinfo {note} {See Supplemental Materials for the details}\BibitemShut
  {NoStop}%
\bibitem [{\citenamefont {Hadad}\ \emph {et~al.}(2018)\citenamefont {Hadad},
  \citenamefont {Soric}, \citenamefont {Khanikaev},\ and\ \citenamefont
  {Al{\`u}}}]{Khanikaev2018}%
  \BibitemOpen
  \bibfield  {author} {\bibinfo {author} {\bibfnamefont {Y.}~\bibnamefont
  {Hadad}}, \bibinfo {author} {\bibfnamefont {J.~C.}\ \bibnamefont {Soric}},
  \bibinfo {author} {\bibfnamefont {A.~B.}\ \bibnamefont {Khanikaev}}, \ and\
  \bibinfo {author} {\bibfnamefont {A.}~\bibnamefont {Al{\`u}}},\ }\bibfield
  {title} {\enquote {\bibinfo {title} {Self-induced topological protection in
  nonlinear circuit arrays},}\ }\href {\doibase 10.1038/s41928-018-0042-z}
  {\bibfield  {journal} {\bibinfo  {journal} {Nature Electronics}\ }\textbf
  {\bibinfo {volume} {1}},\ \bibinfo {pages} {178--182} (\bibinfo {year}
  {2018})}\BibitemShut {NoStop}%
\bibitem [{\citenamefont {Schindler}\ \emph {et~al.}(2018)\citenamefont
  {Schindler}, \citenamefont {Cook}, \citenamefont {Vergniory}, \citenamefont
  {Wang}, \citenamefont {Parkin}, \citenamefont {Bernevig},\ and\ \citenamefont
  {Neupert}}]{Schindler2018}%
  \BibitemOpen
  \bibfield  {author} {\bibinfo {author} {\bibfnamefont {F.}~\bibnamefont
  {Schindler}}, \bibinfo {author} {\bibfnamefont {A.~M.}\ \bibnamefont {Cook}},
  \bibinfo {author} {\bibfnamefont {M.~G.}\ \bibnamefont {Vergniory}}, \bibinfo
  {author} {\bibfnamefont {Z.}~\bibnamefont {Wang}}, \bibinfo {author}
  {\bibfnamefont {S.~S.~P.}\ \bibnamefont {Parkin}}, \bibinfo {author}
  {\bibfnamefont {B.~A.}\ \bibnamefont {Bernevig}}, \ and\ \bibinfo {author}
  {\bibfnamefont {T.}~\bibnamefont {Neupert}},\ }\bibfield  {title} {\enquote
  {\bibinfo {title} {Higher-order topological insulators},}\ }\href {\doibase
  10.1126/sciadv.aat0346} {\bibfield  {journal} {\bibinfo  {journal} {Science
  Advances}\ }\textbf {\bibinfo {volume} {4}} (\bibinfo {year} {2018}),\
  10.1126/sciadv.aat0346}\BibitemShut {NoStop}%
\bibitem [{\citenamefont {Dmitriev}\ and\ \citenamefont
  {Suris}(2002)}]{Dmitriev2002}%
  \BibitemOpen
  \bibfield  {author} {\bibinfo {author} {\bibfnamefont {I.~A.}\ \bibnamefont
  {Dmitriev}}\ and\ \bibinfo {author} {\bibfnamefont {R.~A.}\ \bibnamefont
  {Suris}},\ }\bibfield  {title} {\enquote {\bibinfo {title} {Damping of
  {B}loch oscillations in quantum dot superlattices: {A} general approach},}\
  }\href {\doibase 10.1134/1.1529248} {\bibfield  {journal} {\bibinfo
  {journal} {Semiconductors}\ }\textbf {\bibinfo {volume} {36}},\ \bibinfo
  {pages} {1364--1374} (\bibinfo {year} {2002})}\BibitemShut {NoStop}%
\end{thebibliography}

%merlin.mbs apsrev4-1.bst 2010-07-25 4.21a (PWD, AO, DPC) hacked
%Control: key (0)
%Control: author (8) initials jnrlst
%Control: editor formatted (1) identically to author
%Control: production of article title (0) allowed
%Control: page (1) range
%Control: year (0) verbatim
%Control: production of eprint (0) enabled
%

\end{document}